\documentclass[entropy,article,accept,moreauthors,pdftex,12pt,a4paper]{mdpi} 
\definecolor{myColor}{rgb}{0.4,0.6,0.4}

\lastpage{x}
\doinum{10.3390/------}
\pubvolume{xx}
\pubyear{2021}
\history{Received: xx / Accepted: xx / Published: xx}

\Title{Collective strategy condensation: When envy splits
societies 
}

\Author{Claudius Gros}

\address{%
Institute for Theoretical Physics, Goethe University
Frankfurt, Germany\\
        }

\corres{gros07[@]itp.uni-frankfurt.de}

\abstract{
Human societies are characterized, besides others,
by three constituent features.
(A) Options, as for jobs and societal positions,
differ with respect to their associated monetary and 
non-monetary payoffs.
(B) Competition leads to reduced payoffs when
individuals compete for the same option with others.
(C) People care how they are doing relatively to 
others. The latter trait, the propensity to compare one's own 
success with that of others, expresses itself as envy.
\newline
It is shown that the combination of (A)-(C) leads
to spontaneous class stratification. Societies of agents split 
endogenously into two social classes, an upper and a lower 
class, when envy becomes relevant. A comprehensive analysis 
of the Nash equilibria characterizing a basic reference game is 
presented. Class separation is due to the condensation of
the strategies of lower-class agents, which play an
identical mixed strategy. Upper class agents do not condense,
following individualist pure strategies. 
\newline
Model and results are size-consistent, holding for 
arbitrary large numbers of agents and options. Analytic 
results are confirmed by extensive numerical simulations.
An analogy to interacting confined classical particles
is discussed.
         } 


\begin{document}


\newpage
\section{Introduction}

The notion of an `ideal society' has always
been controversial \cite{reeve2004plato,swidler1991ideal},
especially regarding the conditions for
social classes to arise endogenously when by-birth 
privileges and handicaps are absent, a feature
commonly presumed to be desirable.
In this regard one may consider a society to be
`ideal' when the playing ground is fair, which
means that members have equal access to societal 
options and positions. Here we examine this situation 
using a generalized game theoretical setting.

Two building blocks constitute the core of most abstract 
games \cite{myerson2013game}: competition and that 
different options yield distinct rewards. In this study 
we examine what happens if a third element is 
added, postulating that agents desire to compare 
their individual success reciprocally, a trait 
usually termed `envy' \cite{nguyen2014minimizing}. 
We show that envy splits ideal societies. Two distinct social 
classes, an upper and a lower class, form endogenously
when the desire to compare success becomes substantial.

The notion of envy is based on the observation
that the satisfaction individuals receive from
having and spending money depends not only on the
absolute level of consumption, but also on how
one's own consumption level compares with that of
others \cite{hopkins2004running}. This view, which 
is at the heart of relative income theory
\cite{mcbride2001relative,clark2010compares},
is taken for granted, to give an example, when 
poverty is defined not in absolute, but in relative 
terms \cite{sen1983poor,wagle2002rethinking}.

Key to our research is the notion that class
structures may emerge from class-neutral
interactions between individual agents.
On an equivalent basis, a large body of
social computation research \cite{conte2012manifesto,lane2018cognition}
has investigated to which extent cooperation 
\cite{janssen2010lab,gachter2009reciprocity},
reciprocity \cite{fehr2000fairness},
altruism \cite{bowles2006group} and
social norms \cite{bicchieri2005grammar}
are emergent phenomena. The model investigated in
this study is formulated directly in terms of
strategies, as usual for animal conflict models
\cite{smith1974theory}, like the war of attrition.
A corresponding agent-based simulation setup would
also be possible, with the differences vanishing
in the limit of large numbers of agents and behavioral
options. The adaptive game-theoretical formulation 
used here comes with the advantage that the properties 
of the class stratified state can be studied analytically.
Our study can be seen as a generalization of evolutionary 
game theory, which is dedicated in good part to the origins 
of behavioral traits \cite{newton2018evolutionary}, to the 
emergence of class structures. Other alternatives include
dynamical systems investigations of the stability of societies
\cite{gros2017entrenched}, and game theoretical approaches
centered on selected key societal players \cite{ridley2020game}.


\subsection{Social classes in terms of reward clusters}

We consider a society of $N$ agents, with every
agent able to select from $M$ options. The
payoff function $E_i^\alpha$, for option $i$
and agent $\alpha$, is agent specific, but only
to the extent that it depends explicitly on the strategies
$p_i^\alpha\ge0$. Strategies are normalized, 
$\sum_i p_i^\alpha=1$, with $p_i^\alpha\ge0$ 
denoting the probability that agent $\alpha$ 
selects option $i$. Rewards $R^\alpha$ are defined 
as the expected payoffs,
\begin{equation}
R^\alpha = \sum_i E_i^\alpha p_i^\alpha
=\left\langle E_i^\alpha \right\rangle_{\{\rho_i^\alpha\}}\,.
\label{reward}
\end{equation}
The number of options $M$ can be both larger or 
smaller than the number of agents $N$, with 
size-consistent large-$N$ limits being recovered
for constant ratios $\nu=M/N$. 

\begin{figure}[!t]
\centering
\includegraphics[width=0.65\columnwidth]{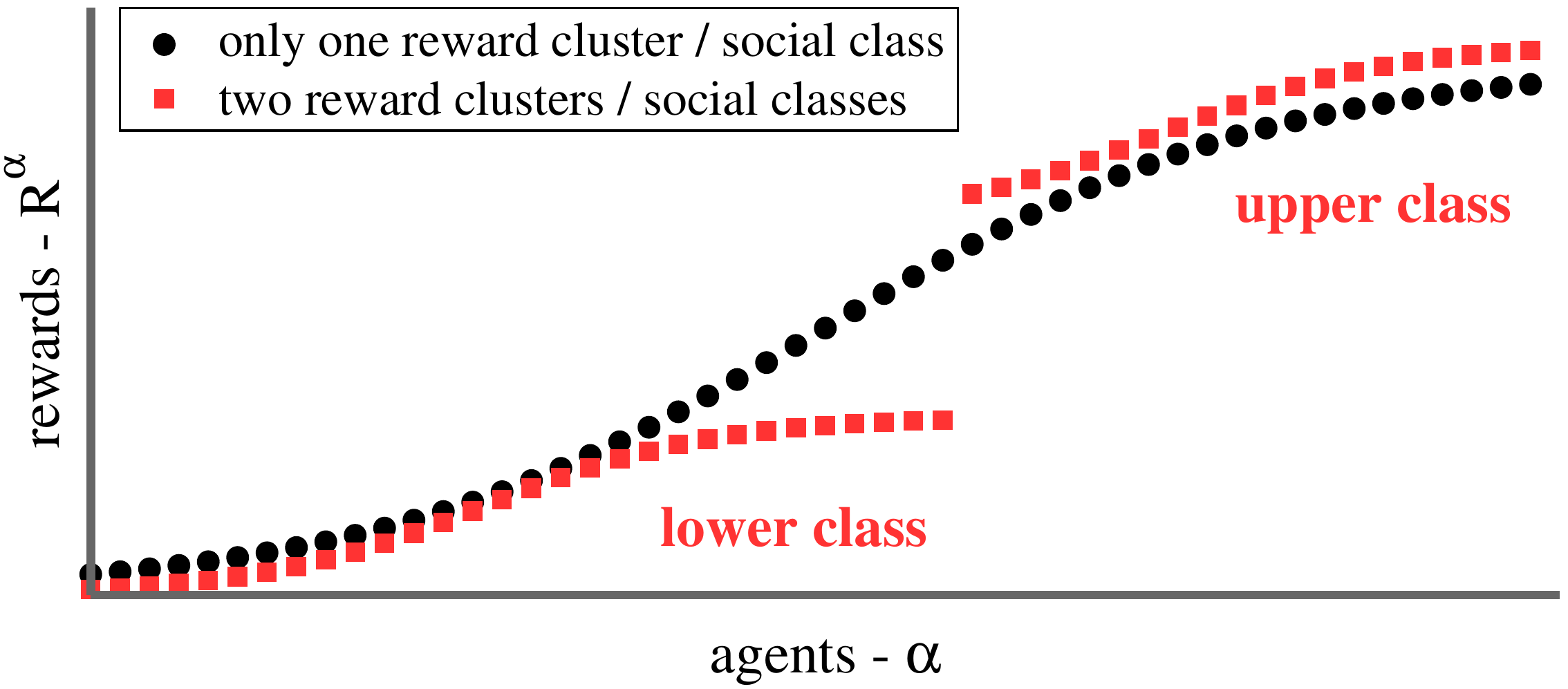}
\caption{{\bf Social classes as reward clusters.}
When ordering the rewards $R^\alpha$ obtained by agents
$\alpha$ as a function of size, the resulting reward spectrum 
may be characterized either by a continuous distribution 
(black), or by one or more gaps (red). Clustering rewards 
with regard to proximity allows consequently for a 
bare-bone definition of social classes. 
}
\label{fig_rewardClusters}
\end{figure}

In this study we define social classes in terms
of reward clusters. Agents within the same
class receive rewards similar in magnitude, which
are separated by a gap from the rewards of
other classes, as illustrated in Fig.~\ref{fig_rewardClusters}. 
In human societies, social groups are shaped in 
particular also by the notion of social identity 
\cite{scheepers2019social}, which is absent
in the bare-bone definition of social classes 
used here. A political theory for social classes
is beyond the scope of the present study.

\section{Shopping trouble model}

We define with
\begin{equation}
\bar{R}=\frac{1}{M}\sum_\alpha R^\alpha\,,
\label{R_bar}
\end{equation}
the mean reward $\bar{R}$ of all agents. 
The payoff function 
\begin{equation}
E_i^\alpha = v_i -\kappa
\sum_{\beta\ne \alpha} p_i^\beta
+\varepsilon\, p_i^\alpha\log\left(\frac{R^\alpha}{\bar{R}}\right)
\label{STM}
\end{equation}
of our reference model contains three terms:
\begin{itemize}
\item {\bf Basic utility.} The basic utility function
      $v_i$, which is identical for all agents, encodes
      the notion that options come with different payoffs.
      Mapping options to qualities $q_i\!\in\![0,2]$, we will
      use a simple inverted parabola, $v_i=1-(1-q_i)^2$,
      for the basic utility.
\item {\bf Competition.} There is a flat penalty $\kappa$
      for agents competing heads on. Payoff reduction is
      proportional to the probability $p_i^\beta$ that
      other agents select the option in question. The
      respective combinatorial factors are approximated
      linearly in (\ref{STM}), as given by the sum
      $\sum_{\beta\ne \alpha} p_i^\beta$.
\item {\bf Envy.} One's own success with respect to the
      mean reward, $R^\alpha/\bar{R}$, induces a psychological 
      reward component.
\end{itemize}
The log-dependency $\log(R^\alpha/\bar{R})$ of the
envy term in (\ref{STM}) reflects the well established
observation that the brain discounts
sensory stimuli \cite{hecht1924visual},
numbers \cite{dehaene2003neural},
time \cite{howard2018memory}, and
data sizes \cite{gros2012neuropsychological}
logarithmically. In addition, the envy term is 
proportional to the current probability 
$p_i^\alpha$ to select option $i$, which encodes
a straightforward rational. When everything is fine, 
when $\log(R^\alpha/\bar{R})\!>\!0$, the current strategy
is reinforced, and suppressed when $\log(R^\alpha/\bar{R})\!<\!0$.
The effect is that agents with high/low rewards tend
to pursuit pure/mixed strategies.
Eq.~(\ref{STM}) is called the `shopping trouble model',
as it can be applied, besides the general social context,
to the case that agents need to optimize their shopping
list \cite{gros2020self}. Note that that agents have only
a single goal within the shopping trouble game, reward 
maximisation, in contrast to most status seeking games 
\cite{congleton1989efficient,haagsma2010equilibrium},
for which both status and utility are separately important
\cite{shi2017nonlinear,courty2019pure}.

\begin{figure}[!t]
\centering
\includegraphics[width=0.45\columnwidth]{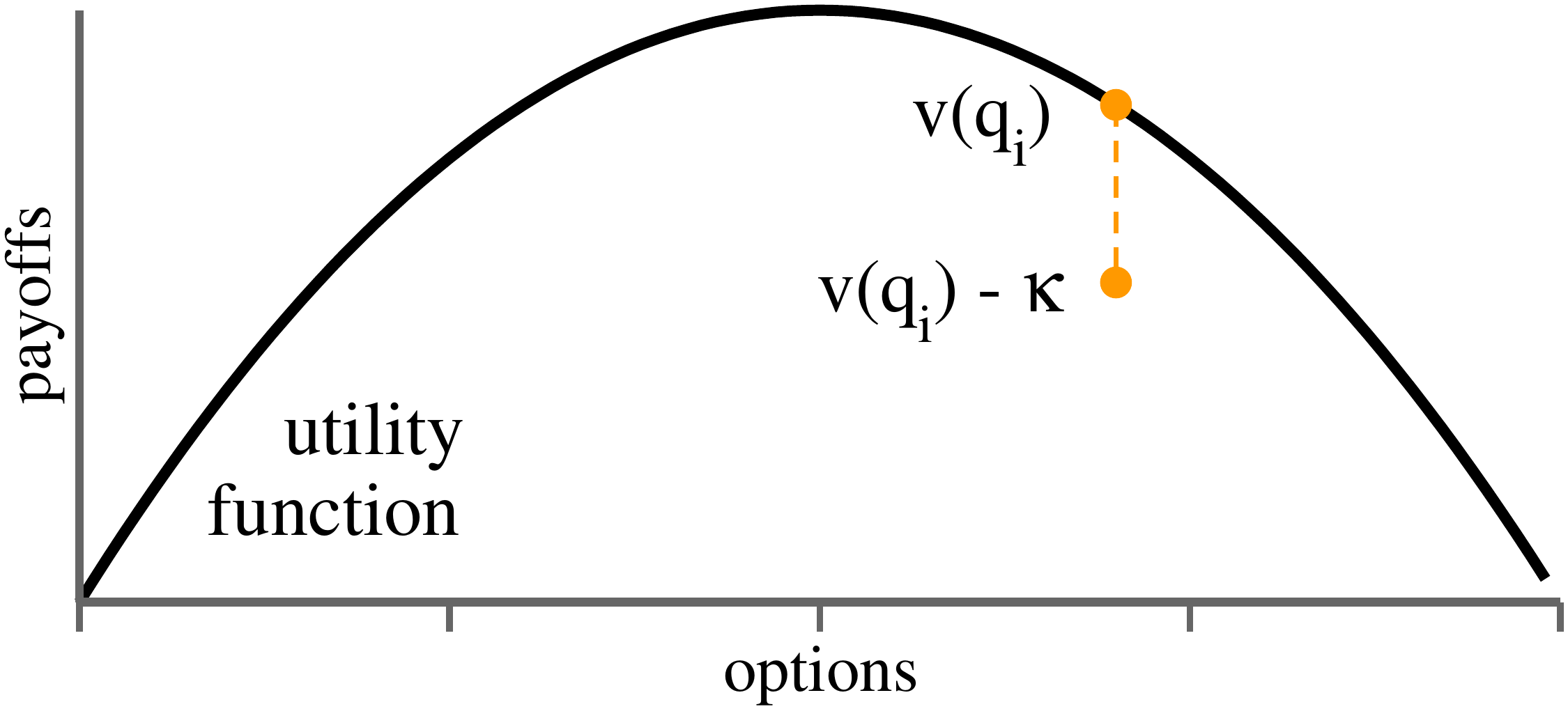}\hfill
\includegraphics[width=0.45\columnwidth]{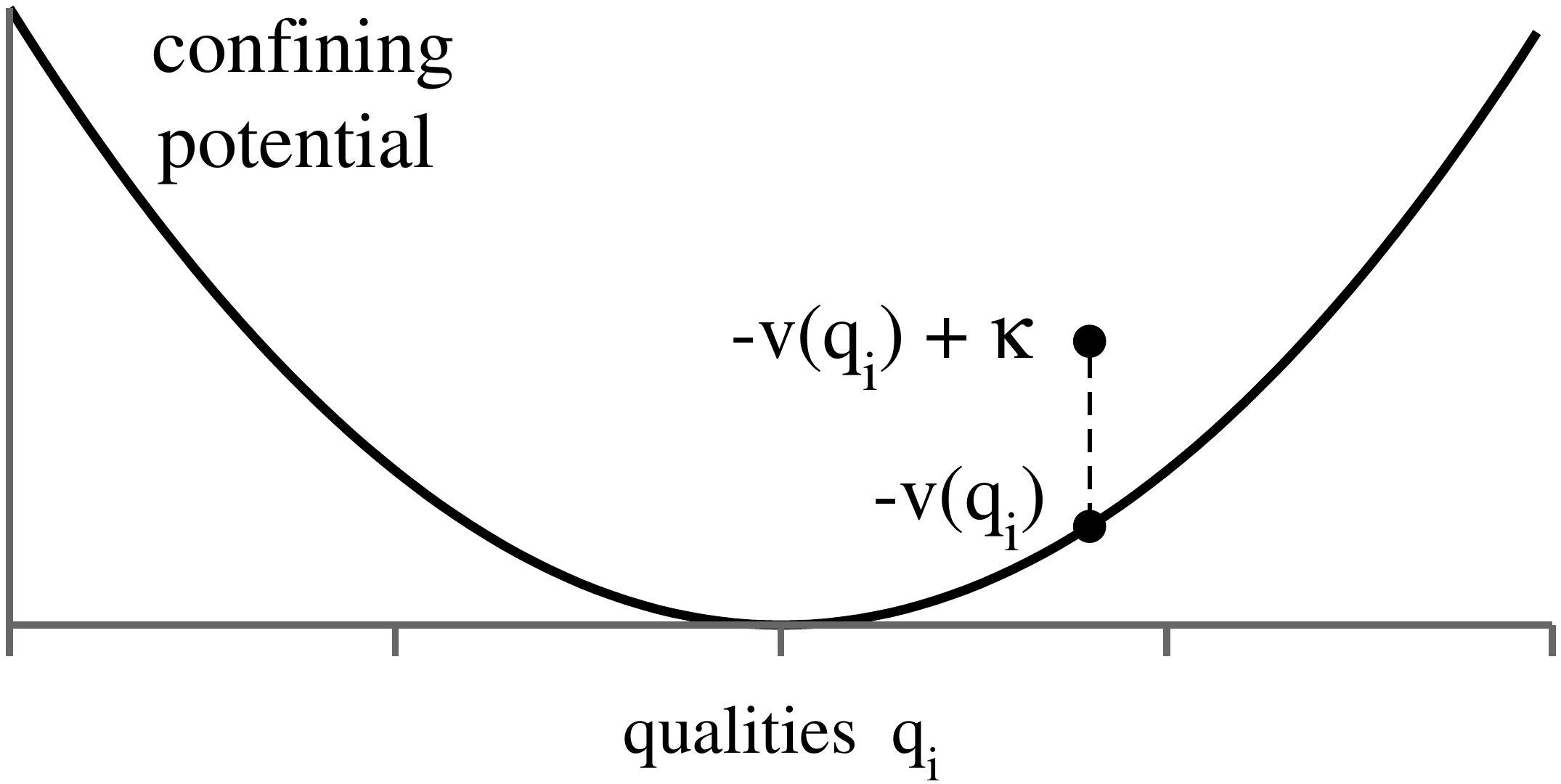}
\caption{{\bf Correspondence to interacting classical particles.}
{\sl Left:} Agents selecting a strategy $i$ receive a bar utility
$v(q_i)$ (inverted parabola), which is reduced by a flat amount
$\kappa$, the competition term, if another agent selects the same option.
Utilities are to be maximized.
\newline
{\sl Right:} Classical particles in a confining potential $-v(q_i)$
(parabola) repel each other by an amount $-\kappa$. Energy is minimized.
}
\label{fig_physicsCorrespondence}
\end{figure}

\subsection{Correspondence to confined interacting classical particles}

The shopping trouble model (\ref{STM}) can be interpreted
in terms of interacting and confined classical particles,
with the correspondence
 
\smallskip
\hfill\begin{tabular}{rcl}
agents            &$\leftrightarrow$& classical particles \\
$q_i$             &$\leftrightarrow$& states \\
$-v(q_i)$         &$\leftrightarrow$& confining potential \\
$\kappa$          &$\leftrightarrow$& Coulomb repulsion\\
$\epsilon$        &$\leftrightarrow$& energy-dependent interaction
\end{tabular}\hfill
\smallskip
 
\noindent
as is illustrated in Fig.~\ref{fig_physicsCorrespondence}.
Without the energy-dependent interaction term, $\epsilon$ (envy), 
particles settle into the respective lowest energy states, which 
are given by $-v(q_i)+(n_i-1)\kappa$, where $n_i$ is the occupation 
number of state $q_i$ (the number of agents selecting the quality 
$q_i$).  Strategies $p_i^\alpha$ correspond in physics terms
to the occupation distribution. At finite temperature particles
can always swap places, which implies that strategies are identical.
This is however not the case at zero temperature. Finite 
temperatures correspond in game theory that agents select 
alternative, lower-reward strategies, with a probability
given by the respective Boltzmann factors. Here we work
with strictly rational, zero-temperature agents, which 
go always for the best choice.

\begin{figure}[!t]
\centering
\includegraphics[width=0.85\columnwidth]{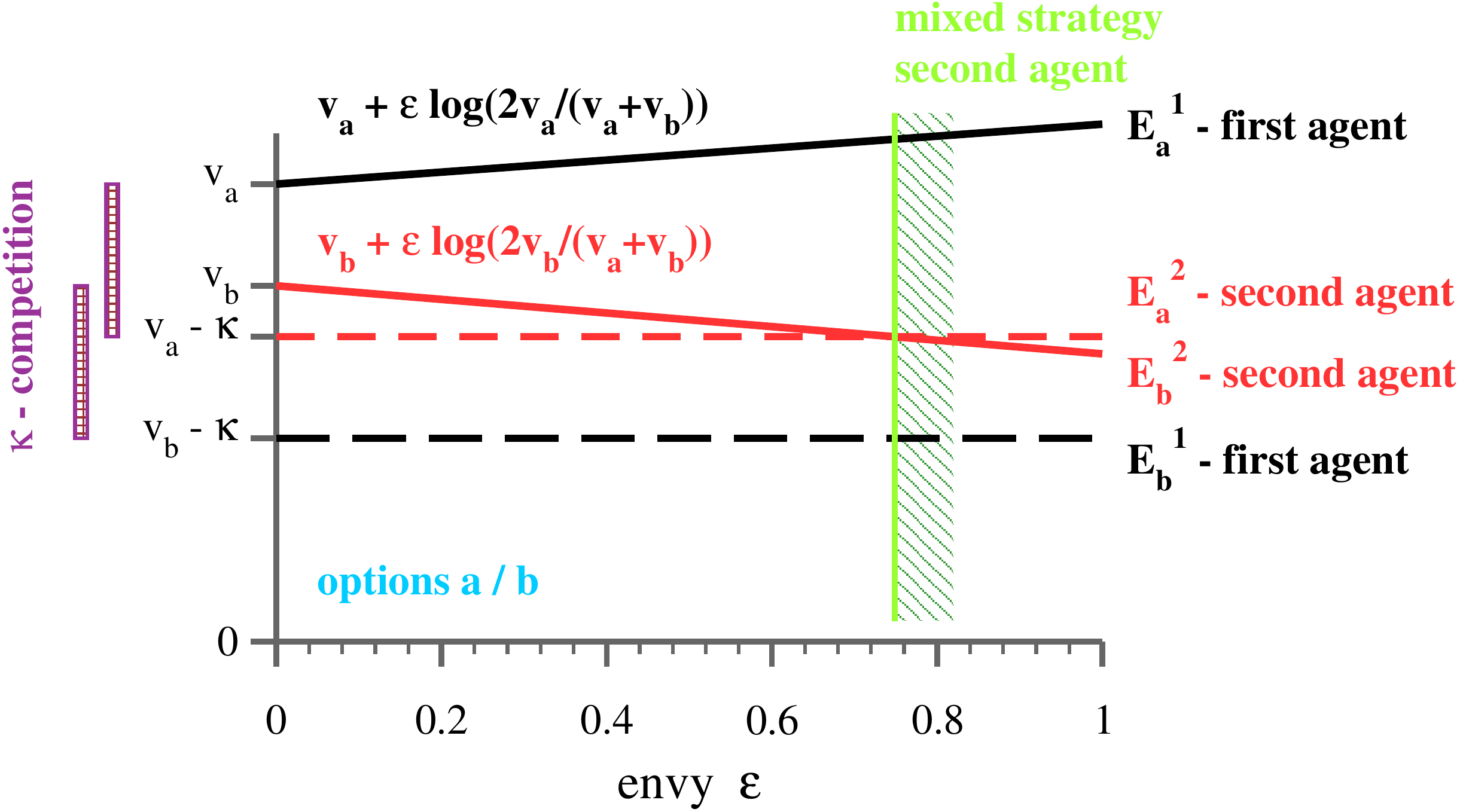}
\caption{{\bf Envy-induced transition from pure to mixed strategies.}
Illustration of the case of two agents that can select between two 
options, $a/b$, with basic utilities $v_a$ and $v_b$. Here $v_a>v_b$. 
In the absence of envy, 
$\epsilon=0$, both agents play pure strategies, here with the 
first/second agent selecting $a/b$. It would be unfavorable for 
the second agent to invade option $a$, as $v_a-\kappa<v_b$, and 
vice versa, where $\kappa$ is the strength of the competition. 
In this state rewards are $R_a=v_a$ and $R_b=v_b$ and
$R_{a,b}/\bar{R}=2v_{a,b}/(v_a+v_b)$. For the second agent
the envy term $\epsilon p_i^\alpha\log(R^\alpha/\bar{R})$
is negative for the $b$-option, vanishing for the $a$-option.
The second agent starts to play a mixed strategy (green shaded area)
when the payoff $E_b^2 = v_b + \epsilon \log(2v_b/(v_a+v_b))$ 
(red solid line) becomes smaller than the $E_a^2=v_a-\kappa$ 
(red dashed line).
}
\label{fig_twoPlayers}
\end{figure}

\subsection{Strategy evolution}

Agents interact in the shopping trouble model via
two averaging fields \cite{hauert2005game}. The first 
coupling term is a scalar quantity, the average reward $\bar{R}$. 
It quantifies the envy term in (\ref{STM}). The second coupling
term, the mean strategy $\bar{p_i} = \sum_{\beta} p_i^\beta/M$,
is in contrast a function of the available options. It 
enters the penalty term via
\begin{equation}
\sum_{\beta\ne \alpha} p_i^\beta = M\bar{p_i} - p_i^\alpha\,.
\label{ST_penalty_alpha}
\end{equation}
Numerically,  the shopping trouble model is solved
using standard evolutionary dynamics \cite{hofbauer2003evolutionary},
\begin{equation}
p_i^\alpha(t+1) = \frac{p_i^\alpha(t)E_i^\alpha(t)}
{\sum_j p_j^\alpha(t)E_j^\alpha(t)}\,.
\label{evolutionaryDynamics}
\end{equation}
In practice, a constant offset $E_0$ is added on the 
right-hand side, which acts as a smoothing factor.

\subsection{Pure vs.\ mixed strategies}

The support of a strategy $p_i^\alpha$ is given by 
the set of options selected with finite probabilities
$p_i^\alpha>0$. The smallest possible support is one, 
the case of a pure strategy, $p_i^\alpha = \delta_{i,k}$. 
Supports larger than one correspond to mixed strategies.
Without envy, viz when $\epsilon=0$, the Nash stable
strategies of the shopping troubling game are all pure.
Agents just compare the payoff options $v_i-\kappa (n_i-1)$
of distinct options, where $n_i$ is the occupation factor,
viz the number of times option $i$ has been selected
by all agents. If not favorable, agents will
avoid occupied options and settle for lower basic
utilities. The situation is illustrated for
two players in Fig.~\ref{fig_twoPlayers}. By
avoiding each other, agents seemingly cooperate,
a state called `forced cooperation' \cite{gros2020self}.

Relative payoff magnitudes change when envy is introduced.
The own option becomes progressively less attractive
when the envy term is negative, which is the case for
agents below $\bar{R}$, see Fig.~\ref{fig_twoPlayers}. 
Eventually the payoff for the own option levels with
that of an occupied option with a higher basic utility
and mixed strategies appear. For larger numbers of agents,
and options, we find that the evolution of mixed strategies 
with increasing levels of envy leads to a class stratified 
state, as discussed further below.

\begin{figure}[!t]
\centering
\includegraphics[width=0.65\columnwidth]{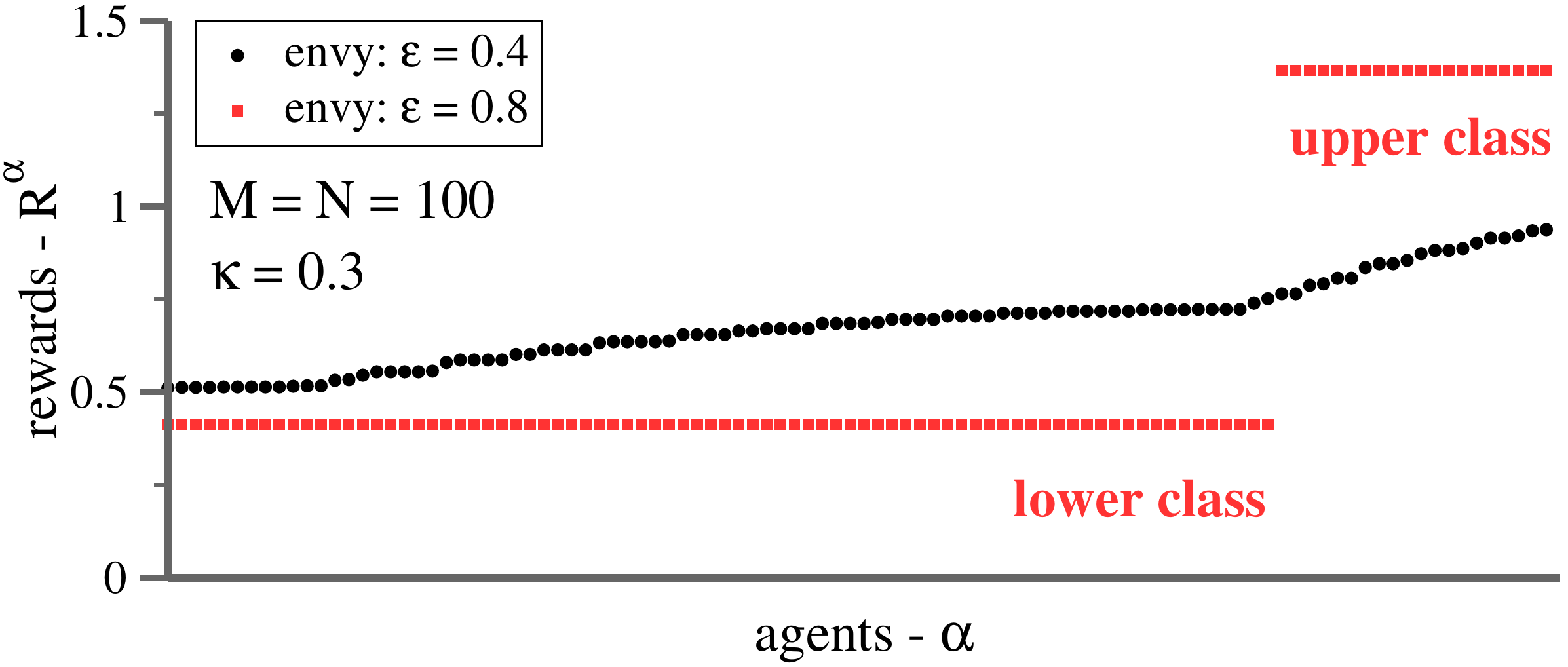}\\
\includegraphics[width=0.65\columnwidth]{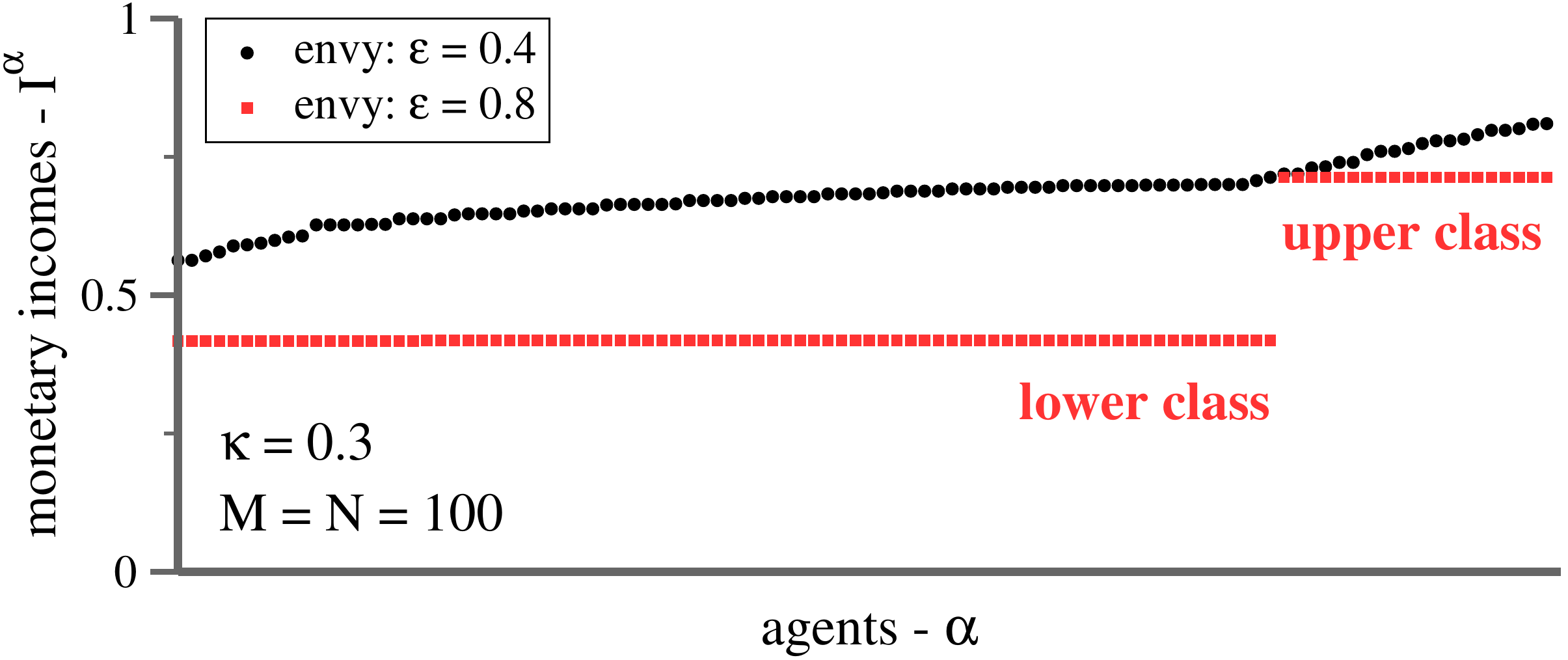}
\caption{{\bf Envy induced class stratification.}
Simulation results for $M=N=100$ and $\kappa=0.3$.
{\sl Top:} For $\epsilon=0.4$ (black) the reward spectrum
is continuous, with agents receiving varying rewards. For
$\epsilon=0.8$ (red) two strictly separated reward clusters 
emerge. Members of the same class receive identical rewards, 
which implies intra-class communism.
{\sl Bottom:} The respective spectrum of monetary incomes
$I^\alpha$, as defined by Eq.~(\ref{income}). The gap 
between lower and upper class is substantial. Note that everybody's
monetary income drops when envy is increased from 0.4
to 0.8. Percentage-wise the loss is comparatively small
for top income agents.
}
\label{fig_rewardIncomeSpectrum}
\end{figure}

\section{Results}

In Fig.~\ref{fig_rewardIncomeSpectrum} representative reward 
distributions for the shopping trouble model are given.
The results are obtained iterating (\ref{evolutionaryDynamics})
recursively for $5\!\cdot\!10^5$ times. The initial strategies
are random, which implies that chance determines the
fate of individual agents, in particular the final
reward. Equivalent results are obtained for smaller and
larger numbers of agents and options. Changing the
density of agents per option, $\nu=M/N$, leads to
quantitative, but not qualitative changes. Larger values
of $\nu$ increase the influence of competition, $\kappa$,
and hence also of envy. The same holds when increasing
$\kappa$ directly.

The transition from forced cooperation at $\kappa=0.4$
to class separation, for $\kappa=0.8$, observed in
Fig.~\ref{fig_rewardIncomeSpectrum} induces a striking
self-organized reorganization of the reward spectrum.
The distribution of rewards is continuous, but otherwise
inconspicuous below the transition. A finite competition
of $\kappa=0.3$ induces cooperation in the sense that
it is in general favorable for agents to select different 
options. Two flat bands arise in contrast in the class
stratified state, one for the upper and one for the
lower class.

\begin{figure}[!t]
\centering
\includegraphics[width=0.65\columnwidth]{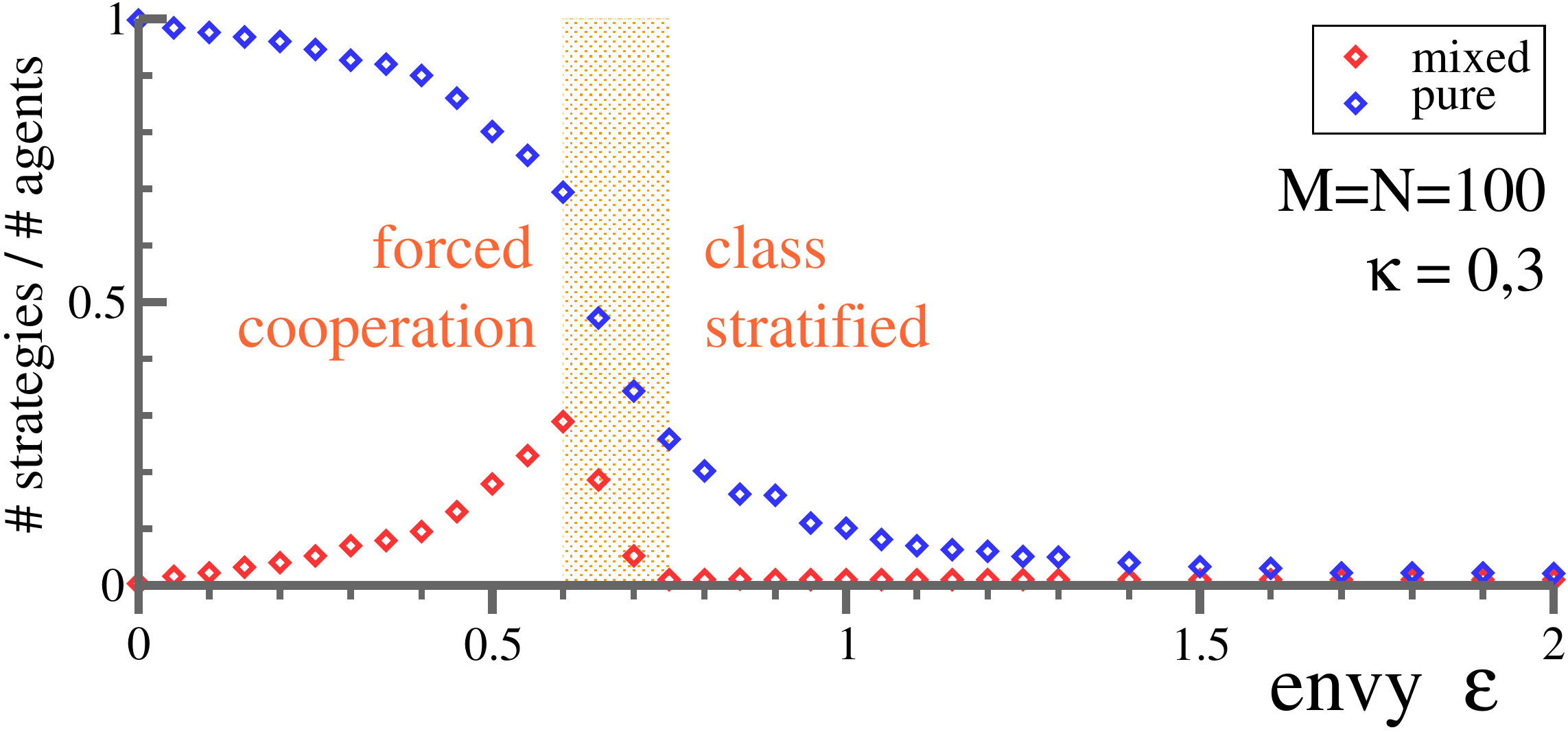}
\caption{{\bf Evolution of mixed strategies.}
For $N\!=\!100$ options and $M\!=\!100$ agents the fraction of
agents playing respectively pure and mixed options.
For small envy the number of mixed strategies raises,
in agreement with the mechanism illustrated in 
Fig.~\ref{fig_twoPlayers} for the case of two players. 
Mixed strategies played by distinct agents merge into a 
single mixed strategy for the entirety of lower-class 
agents once a critical density of mixed strategies is reached. 
The shaded region denotes bistability. When starting from 
random initial strategies and values of $\epsilon$ in the 
shaded region, the evolutionary dynamics (\ref{evolutionaryDynamics}) 
leads to either of two possible Nash equilibria, forced cooperation 
and the class stratification. The fraction of pure strategies 
drops for all $\epsilon$, until only one or two upper class members
remain, the monarchy state. Adapted from \cite{gros2020self}.
}
\label{fig_mixedPure}
\end{figure}

The observation that all lower-class agents receive
identical rewards has a relatively simple explanation.
The number of mixed strategies first raises with
$\epsilon$, in order to drop to one in the class
stratified state. Compare Fig.~\ref{fig_mixedPure}.
Inspecting the individual strategies one by one reveals 
that an identical mixed strategy is played by the entirety
of lower-class agents. Colloquially speaking one
becomes a member of the masses when joining the
lower class. This result explains that a single 
mixed strategy remains in the class stratified
state and that all members of the lower class
receive the same reward.

\begin{figure}[!t]
\centering
\includegraphics[width=0.65\columnwidth]{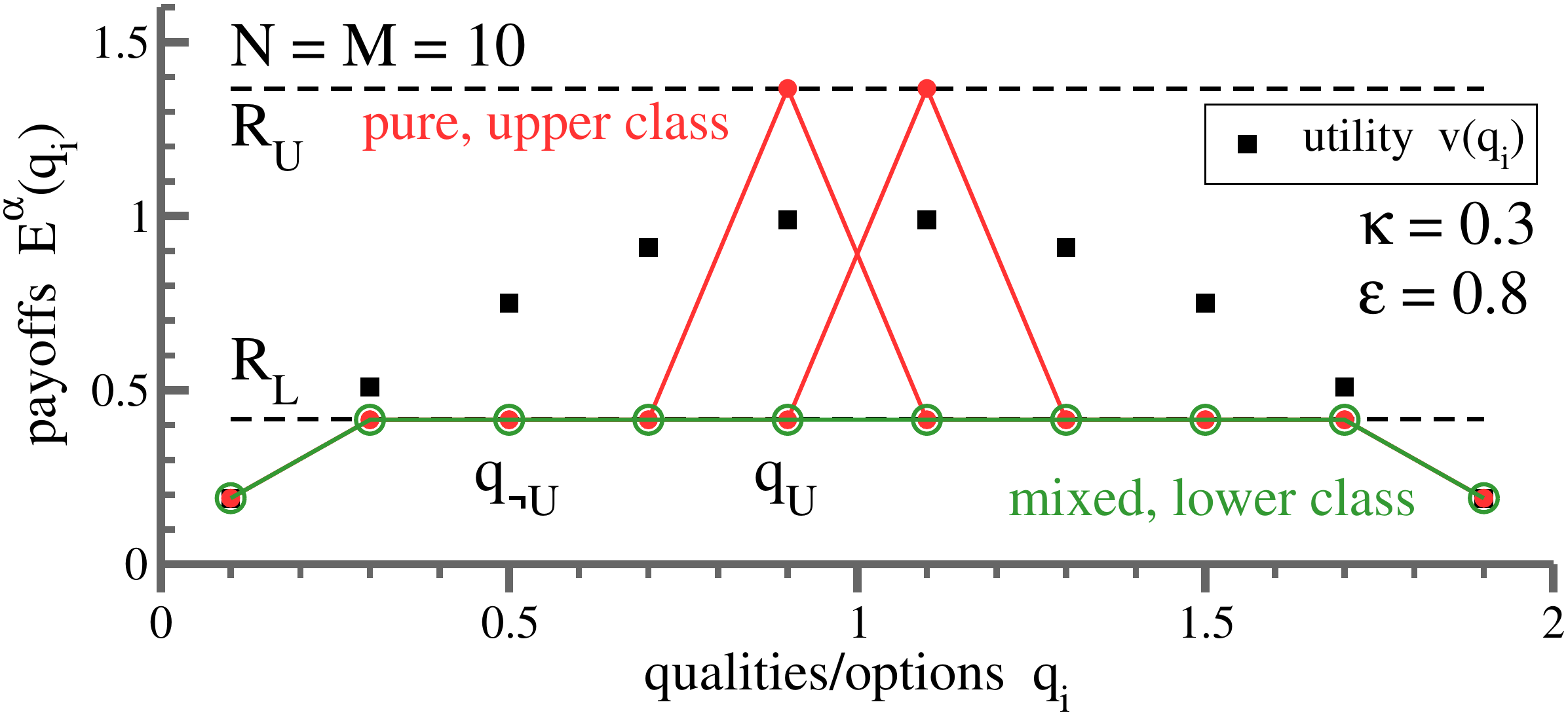}
\caption{{\bf Payoffs in the class stratified state.}
Numerically obtained payoff functions 
$E_i^\alpha=E^\alpha(q_i)$, for a system with 
ten options/agents. The strength of competition/envy
is $\kappa\!=\!0.3$ and $\epsilon\!=\!0.8$. Shown is
the payoff function for the two pure upper-class
strategies (red), and for the single mixed lower-class
strategy (green), played by eight agents. For the
functional form of the bare utility, $v_i=v(q_i)$, an 
inverse parabola has been selected (black squares).
Also shown are the analytic expressions (\ref{R_U})
and (\ref{R_L}) for the upper-/lower-class rewards,
$R_{\rm U}$ and $R_{\rm L}$ (dashed horizontal lines).
Indicated by $q_{\rm U}$ and $q_{\neg\rm U}$ are
qualities played/not played by the upper class.
}
\label{fig_10_10}
\end{figure}

In contrast to the lower class, upper-class agents 
play pure strategies. Members of the upper class
avoid each others, their strategies are hence
individualistic, as illustrated in 
Fig.~\ref{fig_10_10} for a small system.
Why is it then, as evident from
Fig.~\ref{fig_rewardIncomeSpectrum}, that
upper-class agents have identical rewards?
This effect is due to the interaction with
the lower-class mixed strategy, which adapts
itself autonomously, until the contribution from
competition, the term $\sim\!\kappa$ in (\ref{STM}), 
exactly cancels the reward differential arising from
differences in the respective basic utilities
$v_i$. One can trace analytically, as discussed
further below, why this remarkable self-organized
process takes place. 

\subsection{Monetary incomes -- everybody loses}

It is evident from the top panel of
Fig.~\ref{fig_rewardIncomeSpectrum} that
envy induces the formation of two well defined
reward clusters. The question arises, if the gap 
between the lower- and the upper-class cluster is purely 
psychological, viz exclusively due to the envy term
in (\ref{STM}). For this purpose we define with
\begin{equation}
I^\alpha = \sum_i\Big(v_i -\kappa
\sum_{\beta\ne \alpha} p_i^\beta\Big) p_i^\alpha
\label{income}
\end{equation}
the monetary income $I^\alpha$, which represents
the reward $R^\alpha$ minus the envy contribution.
Fig.~\ref{fig_rewardIncomeSpectrum} shows, that
a gap opens both for the reward and for the monetary
income. Everybody loses when envy increases, in the 
sense that monetary incomes drop for all agents, also 
for those at the top, when increasing levels of 
envy force the society to class separate.

\subsection{Analytic properties of the class stratified state}

The agent-to-agent interaction is mediated in the
shopping trouble model by two averaging fields,
$\bar{R}$ and $\bar{p_i}$, as discussed further above.
It can be shown \cite{gros2020self}, that this
property allows to derive analytic expressions for 
the rewards of the lower and of the upper class,
respectively $R_{\rm L}$ and $R_{\rm U}$,
\begin{equation}
R_{\rm L} = \varepsilon\,
\frac{1-f_{\rm L}}{\mathrm{e}^{\kappa/\varepsilon}-1}
\log\left(\frac{\mathrm{e}^{\kappa/\varepsilon}-f_{\rm L}}{1-f_{\rm L}} \right)
\label{R_L}
\end{equation}
and
\begin{equation}
R_{\rm U} = \varepsilon\,
\frac{1-f_{\rm L}\mathrm{e}^{-\kappa/\varepsilon}}{1-\mathrm{e}^{-\kappa/\varepsilon}}
\log\left(\frac{\mathrm{e}^{\kappa/\varepsilon}-f_{\rm L}}{1-f_{\rm L}} \right)\,.
\label{R_U}
\end{equation}
Remarkably, above expressions are not explicitly dependent
on the basic utility $v_i$. The only free parameter 
in (\ref{R_L}) and (\ref{R_U}) is the fraction 
$f_L$ of agents in the lower class, which can be determined 
numerically. For the class stratified state shown in
Fig.~\ref{fig_rewardIncomeSpectrum} one has, as an example,
$f_L=80/100=0.8$, as the number of lower- and upper-class 
agents is respectively 80 and 20. Using $f_L=0.8$ in
(\ref{R_L}) and (\ref{R_U}), the resulting values
for $R_L$ and $R_U$ coincide exactly with the values
obtained numerically.

\subsection{Monarchy vs.\ communism}

The continuous downsizing of the upper class
observable in Fig.~\ref{fig_mixedPure} raises an 
interesting question. Is there a critical envy $\epsilon$, 
beyond which the upper class vanishes altogether? In this
case agents would play exclusively the mixed strategy 
of the former lower class, a telltale characteristics of 
a communist state. Rewards would be equally the same
for everybody. This hypothesis can be tested numerically
by using the extracted lower-class mixed strategy as
the starting strategy for all $M$ agents. Even for large
$\epsilon$, we performed simulations up to $\epsilon=20$, 
the communist state is found to be numerically unstable. 
The system converges without exception to a class-separated 
state containing one or two upper-class agents. Within the 
shopping trouble model communism is unstable against monarchy.

\section{Terminology\label{sect_termiology}}

The notation used throughout this study is summarized 
below. The aim of the compendium below is to provide an 
overview, not complete and detailed definitions.

\smallskip\noindent{\bf Options, qualities \& strategies.}\ 
Options correspond to possible actions, such as
making a purchase in a shop. The numerical 
value associated with option $i$ is the quality $q_i$.
Furthermore we differentiate between option and strategy, 
which is defined here as the probability distribution
function $p_i=p(q_i)$ to pursue a given option.

\smallskip\noindent{\bf Pure vs.\ mixed strategies.}\ 
A strategy is pure when the agent plays the identical 
option at all times, and mixed otherwise, viz when
behavior is variable.

\smallskip\noindent{\bf Evolutionary stable strategies.}\ 
Taking the average payoff received as an indicator for
fitness, a given strategy is evolutionary stable if
every alternative leads to a lower fitness. Evolutionary 
stable strategies are Nash stable.

\smallskip\noindent{\bf Support.}\ 
Strategies are positive definite for all 
options, $p^\alpha(q_i)\ge0$. In reality,
$p^\alpha(q_j)$ is finite only for a subset 
of options, the support of the strategy. 
Strategies are pure/mixed when the size of
the support is one/larger than one.

\smallskip\noindent{\bf Payoff/reward.}\ 
The payoff function is a real-valued function of 
the qualities (options). The mean payoff, as
averaged over the current strategy, is the reward.

\smallskip\noindent{\bf Competitive/cooperative game.}\ 
Parties may coordinate their strategies in cooperative 
games, but not in competitive games. For the shopping
trouble game, voluntary cooperation is not possible.

\smallskip\noindent{\bf Collective effects / phase transition.}\ 
The state of a complex system, like a society of agents, 
may change qualitatively upon changing a parameter, f.i.\
the strength of envy. Such a transition corresponds in physics
terms to a phase transition. Phase transitions are in
general due to collective effects, which means that they
are the result of the interaction between the components, 
here the agents.

\smallskip\noindent{\bf Forced cooperation / class stratification.}\ 
Forced cooperation is present when agents seemingly 
cooperate by avoiding each other, as far as possible.
It is forced when agents optimize in reality just
their individual fitness. Forced cooperation and the
class stratified state are separated by a collective
phase transition.

\smallskip\noindent{\bf Envy.}\ 
Envy is postulated to have opposite effects on
agents with high/low rewards. When their reward is
above the average, agents take this as an indication
that they are doing well and that the best course
of action is to enhance the current strategy. Agents
with below-the-average rewards are motivated in
contrast to search for alternatives, viz to
change the current strategy.

\smallskip\noindent{\bf Monarchy \& communism.}\ 
Monarchy and communism are used throughout this
study exclusively for the labeling of states defined 
by specific constellations of strategies. Secondary
characterizations in terms of political theory are 
not implied. Monarchy is present in a class stratified
society when all but one or two agents belong to the
lower class. All members of the society are part of 
a unique class in communism, with everybody receiving
identical rewards and following the same mixed strategy.

\section{Discussion}

The payoff function of the shopping trouble model 
(\ref{STM}) is not static, as in classical
games, but highly adaptive. The payoff received
when selecting a certain action $i$ depends dynamically
on the strategies of the other agents. At its core
this is typical for social simulations studies \cite{conte2012manifesto},
with the twist that the shopping trouble model is
formulated directly in terms of strategies.
The resulting evolutionary stable strategies are
hence to be determined self-consistently, which
implies that certain aspects of the multi-agent Nash 
solutions may have emergent character \cite{gros2015complex}.
This is indeed observed.

The shopping trouble model studied here incorporates
specific functionalities. We believe, however, that
alternative models based on the same three principles,
payoff diversity, competition, and inter-agent reward 
comparison, would lead to qualitatively similar results.
Further we note that the widely used distinction between 
benign and malicious envy \cite{van2016envy} enters the 
shopping trouble model, albeit indirectly. Benign envy, 
the quest to reach a better outcome by improving oneself, 
can be said to be operative when agents select the best pure 
strategy compatible with everybody else's choice. Malicious envy, 
which aims to pull somebody down from their superior position
\cite{van2016envy}, is functionally operative when
agents start to invade somebody else's zone by switching to mixed
strategies, as shown in Fig.~\ref{fig_twoPlayers}.
In this interpretation, societies are pushed towards class 
stratification by malicious, and not by benign envy.
On a societal level, malicious envy is counterproductive.

\section{Conclusions}

We conclude by recapitulating the driving forces
for the class stratification transition. Agents 
with low rewards are constantly searching for better
options (compare Sect.~\ref{sect_termiology}). 
However, more than one option can be sampled only
when using mixed strategies, which implies that
raising levels of envy induce a corresponding 
larger number of mixed strategies, as observed
in Fig.~\ref{fig_mixedPure}. In the end, a large
number of low-reward agents are trying to explore 
extended ranges of options. At a certain level
of envy their respective mixed strategies collide, 
collapsing at this point into a single encompassing 
strategy for the entire lower class. Class
stratification is hence a result of a spontaneous
condensation of strategies.

In effect, class stratification results from the
constant state of discontent of low-reward agents,
taking place right at the point when their continuing 
search for alternatives runs out of options. High-reward 
agents have in contrast little incentive to do anything
else. In order to keep their privileged position they 
just need to concentrate efforts on what they are doing
best, their current strategies.

\section*{Acknowledgment}

The author thanks Carolin Roskothen, Daniel Lambach 
and Roser Valenti for comments.




\end{document}